\documentclass{ws-ijmpd}

\begin{document}

\markboth{Martin Raue}
{THE EXTRAGALACTIC BACKGROUND LIGHT: LOWER VS UPPER LIMITS}

%
\catchline{}{}{}{}{}
%

\title{THE EXTRAGALACTIC BACKGROUND LIGHT:\\ LOWER VS UPPER LIMITS}

\author{MARTIN RAUE}

\address{Max-Planck-Institut f\"ur Kernphysik, Heidelberg, Germany\\
martin.raue@mpi-hd.mpg.de}

\maketitle

\begin{abstract}
The discovery of distant sources of very high energy (VHE) $\gamma$-rays with hard energy spectra enabled to derive strong upper limits on the density of the extragalactic background light (EBL). These limits are close to the lower limits derived from deep source counts. A recent re-determination of the EBL contribution from resolved sources at $3.6\,\mu$m finds a higher EBL density, which is claimed to be in conflict with the assumptions utilized to derive the EBL upper limits from VHE spectra. Here, it is shown that is possible to recover the canonical $\Gamma\sim1.5$ intrinsic spectra for such a higher EBL density.
\end{abstract}


\section{Introduction}

In the recent years, the discovery of distant sources of VHE $\gamma$-rays with hard energy spectra has successfully been used to constrain the density of the EBL.\cite{aharonian:2006:hess:ebl:nature,mazin:2007a,aharonian:2007:hess:0229} VHE $\gamma$-rays interact with low energy photons from the EBL at ultraviolet (UV) to infrared (IR) wavelengths via the pair-production process and are effectively attenuated.\cite{gould:1967a} With assumptions about the intrinsic spectrum emitted at the source upper limits on the EBL density can be derived. Recent paper\cite{aharonian:2007:hess:0229}, utilizing a maximum hardness of the intrinsic spectrum of $\Gamma<1.5$\footnote{The intrinsic spectrum is described by a power law: $dN/dE \sim E^{-\Gamma}$.}, found an EBL density close to or at the level of the lower limits.

Lower limits on the EBL density can be derived from source counts in the UV to IR.\cite{madau:1996a,fazio:2004a} The conversion from number counts to integrated EBL fluxes is not trivial, since the total flux from the sources has to be measured accurately and the incompleteness of the surveys has to be calculated and corrected for. Other factors include e.g. the cosmic variance and the intrinsic diversity of sources/galaxies.
Recently, Ref.~\refcite{levenson:2008a} (LW08) calculated the total EBL density from galaxies in deep \textit{Spitzer} IR images at $3.6{\mu}m$ applying a Monte Carlo Markov chain simulation, also accounting for the faint outer fringes of galaxies. Their derived values  - $7.6^{+1.0}_{-0.6}$\,nW\,m$^{-2}$\,sr$^{-1}$ and $9.0^{+1.7}_{-0.9}$\,nW\,m$^{-2}$\,sr$^{-1}$, depending on the algorithm used to extract the fluxes (Fig.~\ref{fig:ebl}) - are in excess of the previously derived lower limits\cite{fazio:2004a} and close to the values from direct measurements\cite{hauser:2001a}. The consequences of higher EBL densities - possibly harder intrinsic spectra - have already been discussed e.g. in Ref.~\refcite{aharonian:2006:hess:ebl:nature}. Ref.~\refcite{krennrich:2008a} (KDI08) argued that for the VHE source 1ES\,0229+200 ($z = 0.1396$), 1ES\,1218+304 ($z = 0.182$), 1ES\,1101-232 ($z = 0.186$) such a high EBL density should always lead to intrinsic spectra harder than $\Gamma=1.28\pm0.20$. While the deviation of their result with $\Gamma\sim1.5$ is not large ($1.1\sigma$), we will show here that it is quite possible to construct an EBL density, including the higher density reported in LW08, which results in intrinsic spectra with $\Gamma\sim1.5$.

\section{Method and results}

Currently, there are more than 20 extragalactic VHE sources with measured spectra. For the method to derive EBL upper limit discussed in this paper (maximum hardness) only the most distant source with the hardest spectra are relevant.\cite{mazin:2007a} We will, therefore, only consider the spectra of the blazars 1ES\,1101-232\cite{aharonian:2006:hess:ebl:nature} and 1ES\,0229+200\cite{aharonian:2007:hess:0229}. In addition, we will also show results for 1ES\,1218+304\cite{acciari:2009a:veritas:1218} to facilitate the comparison with the results from KDI08.

\begin{figure}[]
\centering{\includegraphics[width=.63\textwidth]{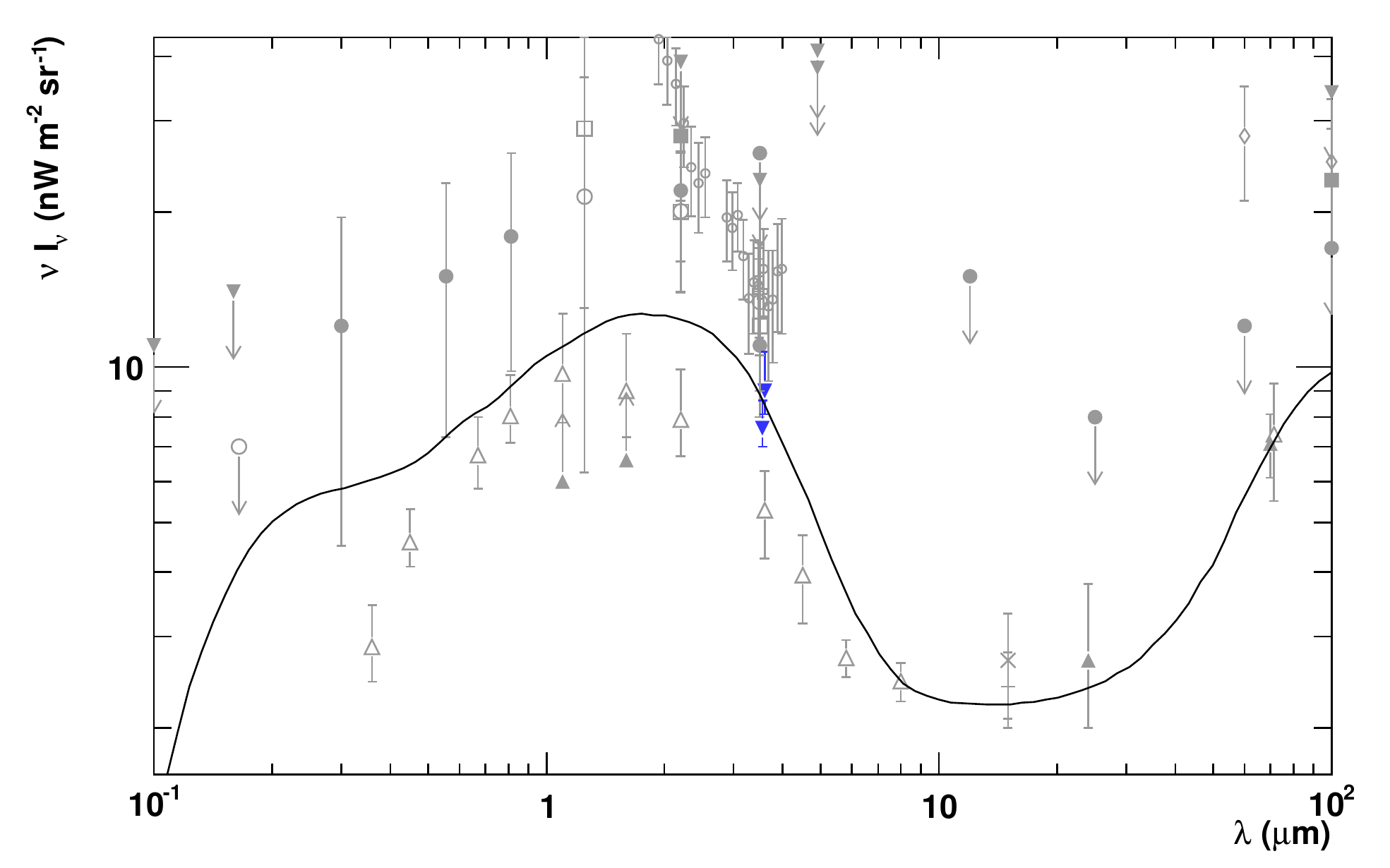}}
\caption{\label{fig:ebl}EBL density utilized in this paper (black solid line). Grey markers: EBL measurements \& limits from Ref.~2; blue markers: results from LW08.}.
\end{figure}

\begin{figure}[]
\centering{\includegraphics[width=0.46\textwidth]{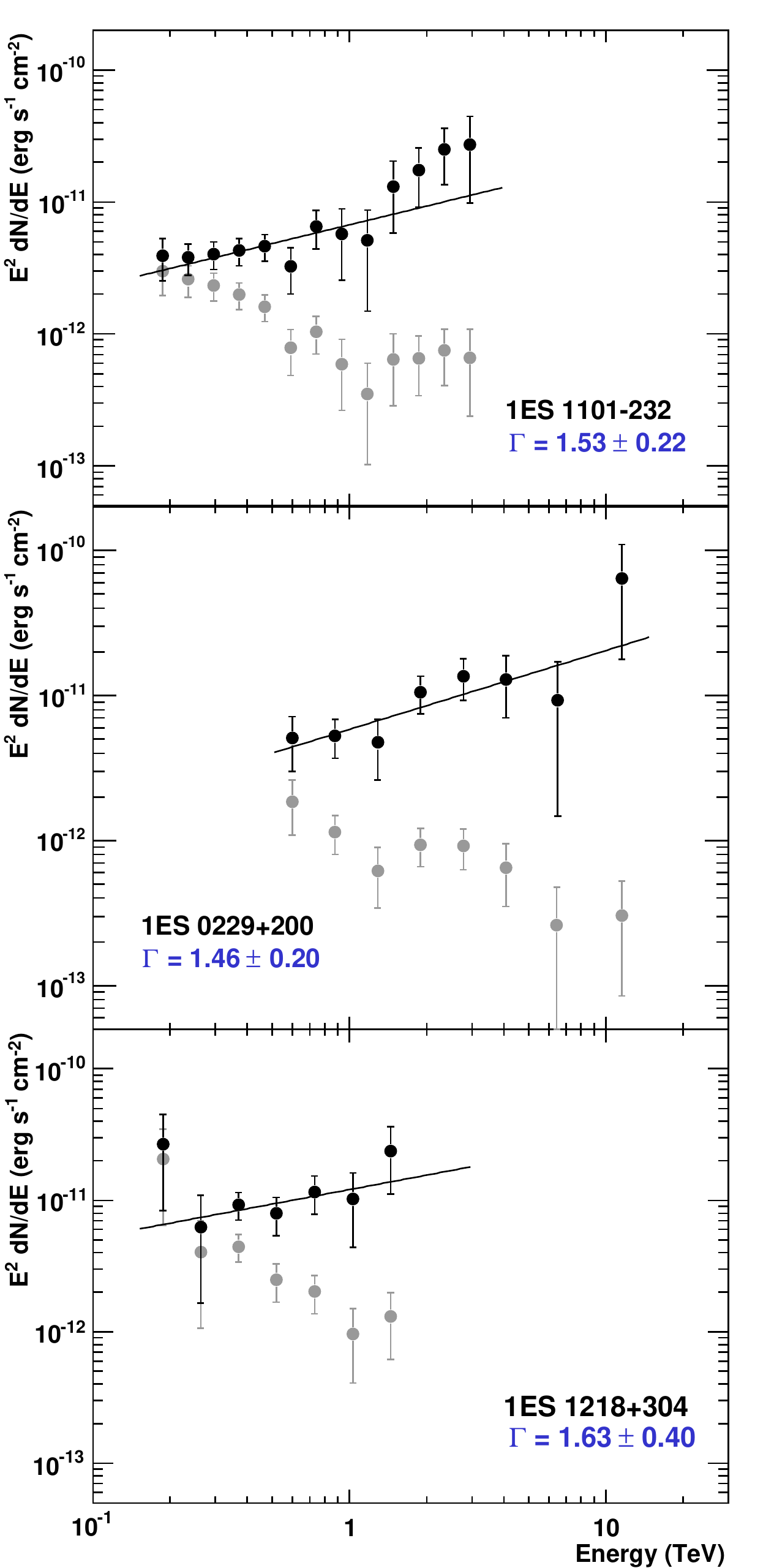}}
\caption{\label{fig:spectra} VHE spectra before (gray) and after correction (black) for the effect from EBL attenuation. The results of a fit of a power law to the intrinsic spectrum are also given.}.
\end{figure}

For the EBL attenuation calculation we follow the method described in Ref.~\refcite{raue:2008b}: an arbitrary generic EBL density at $z=0$ is defined. The EBL evolution is accounted for by adjusting the cosmological density scaling with a factor $f_{evo} = 1.2$: $n(\epsilon)\sim(1+z)^{3 - f_{evo}}$. This choice of $f_{evo}$ has been shown to correctly reproduce the EBL density evolution up to $z\sim0.5$.\cite{raue:2008b}We assume a standard cosmology with $\Omega_{m}=0.3$, $\Omega_\lambda=0.7$ and $h=0.72$. The attenuation per spectrum bin is calculated through a weighted average over the bin size, assuming an intrinsic power-law spectrum with $\Gamma=1.5$. The EBL density used is modified version of the one presented in Ref.~\refcite{raue:2008b} to account for the higher EBL density at $3.6\,{\mu}$m (see Fig.~\ref{fig:ebl}). The resulting intrinsic spectra are shown in Fig.~\ref{fig:spectra} and the fit results are given in Tab.~\ref{ta1}. All intrinsic spectra all well fitted with power laws with $\Gamma\sim1.5$ or larger. Tab.~\ref{ta1} also gives fit results for intrinsic spectra derived by varying parameters of the EBL density or the calculation.

The likely cause of the difference between the results in KDI08 and the ones presented here lies in the tested EBL shapes. As discussed in Ref.~\refcite{mazin:2007a}, the intrinsic spectrum crucially depends on the exact spectral form of the EBL density. Here, especially the slope in the NIR to MIR is important. In addition, accounting for the EBL density evolution result in a change in $\Gamma$ of the order of $0.2$ for $z=0.2$.
The utilized EBL density is low in the mid-IR around $10\,\mu$m, but still compatible with the lower limits. It would be interesting to repeat the analysis of LW08 at these wavelengths.

\begin{table}[]
\tbl{Photon index from the fit of a power-law to the intrinsic spectra for the fiducial EBL density and variations ($h$: Hubble constant, $s$: scaling factor of the EBL density,$f$: generic EBL evolution factor $f_{evo}$, $e$: shift of the energy scale).}
{\begin{tabular}{lccc} \toprule
& 1ES 0229+200 & 1ES 1101-232 & 1ES 1218+304 \\ \colrule
Fiducial & $1.46\pm0.20$ & $1.53\pm0.22$ & $1.63\pm0.40$ \\ \colrule
$h=0.70$ & $1.43\pm0.20$ & $1.49\pm0.22$ & $1.59\pm0.40$ \\
$h=0.74$ & $1.48\pm0.20$ & $1.56\pm0.22$ & $1.68\pm0.40$ \\ \colrule
$s=0.9$ & $1.56\pm0.20$ & $1.67\pm0.22$ & $1.78\pm0.39$ \\
$s=1.1$ & $1.36\pm0.20$ & $1.39\pm0.23$ & $1.49\pm0.41$ \\ \colrule
$f=0.0$ & $1.37\pm0.20$ & $1.37\pm0.23$ & $1.46\pm0.41$ \\
$f=1.0$ & $1.44\pm0.20$ & $1.50\pm0.22$ & $1.61\pm0.40$ \\
$f=1.4$ & $1.47\pm0.20$ & $1.55\pm0.22$ & $1.66\pm0.40$ \\ \colrule
$e=0.9$ & $1.46\pm0.19$ & $1.64\pm0.24$ & $ 1.73\pm0.41$ \\
$e=1.1$ & $1.45\pm0.21$ & $1.45\pm0.21$ & $ 1.57\pm0.38$ \\ \botrule
\end{tabular} \label{ta1}}
\end{table}

\section{Conclusion}

In this paper we discussed the implication of the recent analysis of LW08, reporting a higher EBL density at $3.6\,\mu$m, on the intrinsic VHE spectra of distant sources. We find it is still possible to construct an EBL density which results in intrinsic VHE spectra with $\Gamma\sim1.5$ or larger, contrary to some recent conclusions (KWI08). While the disagreement is not large, the likely origin is the utilized EBL shape. This demonstrates the need of testing a large amount of possible EBL density realization to derive limits on the EBL density. Further analysis of the EBL contribution from resolved sources at other wavelengths would be highly interesting.

\section*{Acknowledgments}
{\small
M.~Raue would like to thank D.~Mazin for many fruitful discussions and F.~Rieger for the careful reading of this manuscript.
This research has been supported by an LEA fellowship and
has made use of NASA's Astrophysics Data System.
}


\def\Journal#1#2#3#4{{#4}, {#1}, {#2}, #3}
\def\NAT{Nature}
\def\AAA{A\&A}
\def\ApJ{ApJ}
\def\AJ{Astronom. Journal}
\def\Aph{Astropart. Phys.}
\def\ApJS{ApJSS}
\def\MNRAS{MNRAS}
\def\NIM{Nucl. Instrum. Methods}
\def\NIMA{Nucl. Instrum. Methods A}
\def\aj{AJ}%
\def\actaa{Acta Astron.}%
\def\araa{ARA\&A}%
\def\apj{ApJ}%
\def\apjl{ApJ}%
\def\apjs{ApJS}%
\def\ao{Appl.~Opt.}%
\def\apss{Ap\&SS}%
\def\aap{A\&A}%
\def\aapr{A\&A~Rev.}%
\def\aaps{A\&AS}%
\def\azh{AZh}%
\def\baas{BAAS}%
\def\bac{Bull. astr. Inst. Czechosl.}%
\def\caa{Chinese Astron. Astrophys.}%
\def\cjaa{Chinese J. Astron. Astrophys.}%
\def\icarus{Icarus}%
\def\jcap{J. Cosmology Astropart. Phys.}%
\def\jrasc{JRASC}%
\def\mnras{MNRAS}%
\def\memras{MmRAS}%
\def\na{New A}%
\def\nar{New A Rev.}%
\def\pasa{PASA}%
\def\pra{Phys.~Rev.~A}%
\def\prb{Phys.~Rev.~B}%
\def\prc{Phys.~Rev.~C}%
\def\prd{Phys.~Rev.~D}%
\def\pre{Phys.~Rev.~E}%
\def\prl{Phys.~Rev.~Lett.}%
\def\pasp{PASP}%
\def\pasj{PASJ}%
\def\qjras{QJRAS}%
\def\rmxaa{Rev. Mexicana Astron. Astrofis.}%
\def\skytel{S\&T}%
\def\solphys{Sol.~Phys.}%
\def\sovast{Soviet~Ast.}%
\def\ssr{Space~Sci.~Rev.}%
\def\zap{ZAp}%
\def\nat{Nature}%
\def\iaucirc{IAU~Circ.}%
\def\aplett{Astrophys.~Lett.}%
\def\apspr{Astrophys.~Space~Phys.~Res.}%
\def\bain{Bull.~Astron.~Inst.~Netherlands}%
\def\fcp{Fund.~Cosmic~Phys.}%
\def\gca{Geochim.~Cosmochim.~Acta}%
\def\grl{Geophys.~Res.~Lett.}%
\def\jcp{J.~Chem.~Phys.}%
\def\jgr{J.~Geophys.~Res.}%
\def\jqsrt{J.~Quant.~Spec.~Radiat.~Transf.}%
\def\memsai{Mem.~Soc.~Astron.~Italiana}%
\def\nphysa{Nucl.~Phys.~A}%
\def\physrep{Phys.~Rep.}%
\def\physscr{Phys.~Scr}%
\def\planss{Planet.~Space~Sci.}%
\def\procspie{Proc.~SPIE}%

\end{document}